\documentclass[aps,prl,amsmath,amssymb,superscriptaddress,showpacs,balancelastpage,10pt,reprint]{revtex4-1}

\usepackage[dvips]{graphicx}
\usepackage{epsfig}
\usepackage[colorlinks=true]{hyperref}
\usepackage{pstricks}
\usepackage{epstopdf}
\usepackage{ulem}

\newcommand{\bket}[1]{\langle #1 \rangle}
\newcommand{\ket}[1]{| #1 \rangle}

\newcommand{\hrho}{\overline{\rho}}
\newcommand{\PP}[1]{{\text{\normalsize$\mathbb P$}}\left(#1\right)}

\newcommand{\tr}[1]{\text{Tr}\left(#1\right)}
\newcommand{\trr}[1]{\text{Tr}^2\left(#1\right)}

\newcommand{\bK}{\boldsymbol{K}}
\newcommand{\ba}{\boldsymbol{a}}
\newcommand{\DD}{\mathcal{D}}

\newcommand{\VV}{\mathcal{V}}

\newcommand{\bR}{\boldsymbol{R}}

\newcommand{\hf}{\overline{f}}

\newcommand{\ML}{\text{\tiny$\!M\!L$}}
\newcommand{\BM}{\text{\tiny$\!B\!M$}}

\newcommand{\AS}[1]{{#1}}

\begin{document}

\title{Quantum state tomography with non-instantaneous measurements, imperfections and decoherence}

\author{P. Six  }
\affiliation{Centre Automatique et Syst\`{e}mes, Mines-ParisTech, PSL Reseach University, 60, bd Saint-Michel, 75006 Paris, France.}
\author{Ph. Campagne-Ibarcq}
\affiliation{Laboratoire Pierre Aigrain, Ecole Normale Sup\'erieure-PSL Research University, CNRS, Universit\'e Pierre et Marie Curie-Sorbonne Universit\'es, Universit\'e Paris Diderot-Sorbonne Paris Cit\'e, 24 rue Lhomond, 75231 Paris Cedex 05, France}
\author{I. Dotsenko}
\affiliation{ Laboratoire Kastler-Brossel, Ecole Normale Sup\'erieure-PSL Research University, Universit\'e Pierre et Marie Curie-Sorbonne Universit\'es, CNRS,
Coll\`{e}ge de France, 11 place Marcelin Berthelot, 75005 Paris, France}
\author{A. Sarlette}
\affiliation{INRIA Paris-Rocquencourt, Domaine de Voluceau, B.P. 105, 78153 Le Chesnay Cedex, France}
\author{B. Huard}
\affiliation{Laboratoire Pierre Aigrain, Ecole Normale Sup\'erieure-PSL Research University, CNRS, Universit\'e Pierre et Marie Curie-Sorbonne Universit\'es, Universit\'e Paris Diderot-Sorbonne Paris Cit\'e, 24 rue Lhomond, 75231 Paris Cedex 05, France}
\author{P. Rouchon }
\email{pierre.rouchon@mines-paristech.fr}
\affiliation{Centre Automatique et Syst\`{e}mes, Mines-ParisTech, PSL Reseach University, 60, bd Saint-Michel, 75006 Paris, France.}

\date{November 27, 2015}


\begin{abstract}
Tomography of a quantum state is usually based on positive operator-valued measure (POVM) and on their experimental  statistics.  Among the available reconstructions,  the  maximum-likelihood (MaxLike)  technique  is an efficient one. We propose an extension of this technique when the measurement process cannot be simply described by an instantaneous POVM. Instead, the tomography relies on a set of quantum trajectories and their measurement records.  This model includes  the fact that, in practice, each measurement could be corrupted by imperfections and decoherence, and could also be associated with the record of continuous-time signals over a finite amount of time.
\AS{The goal is then to retrieve the quantum state that was present at the start of this measurement process.} The proposed extension relies on an explicit expression of the likelihood function via the effective matrices appearing in quantum smoothing and solutions of the adjoint quantum filter. \AS{It allows to retrieve the initial quantum state as in standard MaxLike tomography, but where the traditional POVM operators are replaced by more general ones that depend on the measurement record of each trajectory.} It also provides, aside the MaxLike estimate of the quantum state,  confidence intervals for any observable. Such confidence intervals are derived, as the MaxLike estimate,  from an asymptotic expansion of multi-dimensional Laplace integrals  appearing in  Bayesian Mean estimation. A validation is performed on two sets of experimental data: photon(s) trapped in a microwave cavity subject to quantum  non-demolition measurements relying on Rydberg atoms; heterodyne fluorescence measurements of a  superconducting qubit.
\end{abstract}


\maketitle

\section{Introduction}

Determining efficiently the state of a system whose preparation is imperfectly known is instrumental to quantum physics experiments. Contrarily to classical physics, the determination of a quantum state $\hrho$, its tomography, requires a large number $N$ of independent measurements~\cite{paris-rehacek-book2004}\footnote{See also the lectures at Coll\`{e}ge de France of Serge Haroche {\em Cours 2009-2010: synth\`{e}se et reconstructions d'\'{e}tats quantiques} (in French).}. The state of a quantum system is indeed a statistical quantity by essence, as it encodes the statistics of outcomes for any upcoming measurement. These measurements are usually  modeled using a positive operator-valued measure  (POVM) defined by non-negative self-adjoint operators $\pi_n$ such that $\sum_n \pi_n=I$. The  probability of measurement outcome $n$ is then  given by $\tr{\hrho \pi_n}$. For $N$ large enough,  the reconstruction of $\hrho$ is based on the fact that $\tr{\hrho \pi_n}$ should be close to $N_n/N$ with  $N_n$ the number of  outcomes  $n$ among the $N$ independent measurements: $\sum_n N_n=N$.  Several  reconstruction methods are available. Maximum entropy \cite{buzek:LN04} and compressed sensing \cite{GrossLFBE2010PRL} methods   are well adapted to informationally incomplete sets of measurements. For informationally  complete sets of measurements, maximum likelihood (MaxLike) reconstruction \cite{LvovsR2009RMP} is usually used: it consists in taking   as estimate for $\hrho$ the value  $\rho_{\ML}$ that maximizes the likelihood function denoted by $ \PP{Y~|~\rho}$,  the probability of the measurement data $Y\equiv (N_n)_n$ knowing $\rho$. For the  POVM $(\pi_n)_n$, the likelihood function is directly given by:
\begin{equation}\label{eq:MaxLikePOVM}
\PP{Y~|~\rho}=\prod_{n} \big(\tr{\rho\pi_n}\big)^{N_n}.
\end{equation}
In such usual setting, the measurement process is assumed to be instantaneous  and free from  imperfections and decoherence. In specific experimental situations, such as in~\cite{OurjoJTG2007N}, it is not very difficult to take into account  some  measurement imperfections. In general, the derivation of the likelihood function in presence of imperfections and decoherence during the measurement has not been investigated. This is precisely one of the goals of this paper.   Since  the seminal work of Belavkin, one knows how to take into account measurement imperfection and decoherence for quantum filtering~\cite{Belavkin1992}.   We show here how to exploit the stochastic master equation governing $\rho_t$, the conditional-state  of the  quantum filter, to derive a general expression of the likelihood function: this expression, given by~\eqref{eq:MaxLikeEn},  is a direct generalization of the  above one where the $\pi_n$ are replaced by the adjoint states at the initial times of the $N$ measurement sequences.  These adjoint states obey  a backward master equation appearing  in quantum smoothing~
\cite{Tsang2009PRL,GuevaraWiseman2015} and correspond to the effective operator $E$ defining the past quantum state $(\rho,E)$ introduced in~\cite{GammeJM2013PRL}.

When the support of the likelihood function is mainly concentrated around its maximum at $\rho_{\ML}$, it is known that $\rho_{\ML}$ is a good approximation of the Bayesian Mean estimate $\rho_{\BM}$ defined by (see, e.g.,\cite{Blume2010NJoP}):
\begin{equation}\label{eq:rhoBM}
\rho_{\BM} = \frac{\int_{\DD} \rho~ \PP{Y~|~\rho}\mathbb{P}_0(\rho) d\rho}{\int_{\DD} \PP{Y~|~\rho} \mathbb{P}_0(\rho) d\rho}
\end{equation}
where $\DD$ is the convex set of density operators (here the underlying Hilbert space is of finite dimension) and $\mathbb{P}_{0}(\rho)$ is some prior probability law  of $\rho$ (e.g., Gaussian unitary ensemble~\cite{MehtaBook2004}).  Such approximation of $\rho_{\BM}$  by $\rho_{\ML}$ relies on the first terms of an  asymptotic expansion of multidimensional Laplace integrals~\cite{BleistenHandelsmanBook}  under some regularity conditions.

We show here how such asymptotic expansion provides also a confidence interval for $\tr{\rho_\ML A}$, where $A$ is any Hermitian operator.  Such confidence interval is based on a similar  approximation for  the Bayesian  variance of $\tr{\rho_\ML A}$, denoted by $\sigma^2_\ML(A)$. We provide in~\eqref{eq:Avariance} an  explicit expression that  depends only on  the first and  second-order  derivatives of the log-likelihood function at its maximum $\rho_{\ML}$. When $\rho_{\ML}$  has full rank, we recover the usual asymptotics of  MaxLike estimators  involving  the Fisher information matrix and the Cram\'er-Rao bound~(see, e.g., \cite{MoulinesBook2005}). When $\rho_{\ML}$ is rank deficient, the likelihood reaches its maximum  on the boundary of $\DD$. In this case, such explicit expression approximating the Bayesian variance is not usual and, as far as we know, seems to be new.
\AS{From a practical viewpoint, our MaxLike estimator thus provides a statistically efficient reconstruction method -- exploiting all measurements and information about the dynamics of the measurement process -- along with a confidence bound about the estimate, provided the measurement model is correct. In principle, an uncertain parameter in the dynamics of the measurement model could also be MaxLike estimated by looking at the likelihood of the measurement data conditioned on the parameter value, although this is not covered in the present paper. Some robustness to model errors is illustrated in the experimental validation section.}

After developing the general theory, we report such quantum-state reconstructions for two different sets of  experimental data. The first set corresponds to the quantum non-demolition photon counting in a cavity using Rydberg atoms, and presented in~\cite{PastQuantumStateLKB2014}. The stochastic master equation is a discrete-time Markov chain whose state is the photon-number population. The second set corresponds to the fluorescence heterodyne measurements of a superconducting  qubit  presented in~\cite{CampagneEtAlPRL2014},  a system described by a continuous-time stochastic master equation driven by two Wiener processes. In both cases, we compute the MaxLike estimates $\rho_\ML$ \AS{of $\hrho$, the initial state whose tomography we are supposed to make.  For any time  $t$ between $0$ and $T$,  we can ignore  the  measurement outcomes between  $0$ and $t$ and just retain the measurement outcomes between  $t$ and $T$ for the tomography. Thus we can artificially investigate the result of such tomography on the quantum state  at time $t$, namely the state that would result from decoherence between $0$ and~$t$. } We also give, for physically interesting observable $A$,  the usual 95\% confidence interval via the approximation  $\tr{\rho_\ML A}\pm 2\sigma_\ML(A)$. This confidence interval just means that, if we  perform another tomography with another similar data-set, the probability that the MaxLike estimation of $\tr{\hrho A}$ remains  between  $\tr{\rho_\ML A}-2\sigma_\ML(A)$ and $\tr{\rho_\ML A}+ 2\sigma_\ML(A)$ is greater than 95\%.
This probability has nothing to do with the quantum stochastic character of $\hrho$. Here, $\hrho$ is considered, from a classical statistical point of view, as an unknown constant parameter of a probability law. \AS{More generally, any constant parameter appearing  in  the quantum filter  governing the conditional-state  can be estimated in a similar way, see, e.g., \cite{SixCDC2015}.}

The following section is devoted to discrete-time systems. First, we show how to obtain, from the discrete-time formulation of the quantum filter, an explicit expression of the likelihood function; then, we give the expression of the confidence interval and use it on the first set of experimental data related to the detection of a photon creation quantum jump. In another section, we show how to apply the discrete-time formulation to continuous-time stochastic master equation driven by  Wiener processes in order to obtain a numerical algorithm for computing the adjoint states and the likelihood function. Then, we use this numerical algorithm to estimate the initial state of a qubit relaxing towards its ground state and submitted to the heterodyne measurement of its  fluorescence. In appendix, we give the main calculations yielding the asymptotic expression of the Bayesian variance for any observable $A$.

\section{Discrete-time setting} \label{sec:discrete}

 \subsection{Quantum filtering}

We have at our disposal $N$  realizations, starting from the same initial state $\hrho$ that we want to determine,   and producing  $N$ measurement records  $(y_t^{(n)})_{t = 0, \ldots, T_n}$,  indexed by $n\in\{1,\ldots,N\}$ and where the time $t$  corresponds to an integer between $0$ and $T_n$, the duration  of realization $n$. For each  realization,  the quantum filter  provides the conditional state at $t$, denoted  by  $\rho_t^{(n)}$, conditioned on $\rho_0^{(n)}=\rho$ and knowing the past measurements  $(y_0^{(n)},\ldots, y_{t-1}^{(n)})$. This filter, a discrete-time version of Belavkin's continuous-time filter,  is defined by a family of completely positive maps $\bK_{y,t}$ indexed by the time $t$ and the measurement outcome $y$ (see, e.g., \cite{dotsenko-et-al:PRA09,wiseman-milburnBook,somaraju-et-al:acc2012}). Moreover, for each $t$,  the  completely positive map $\bK_t=\sum_{y} \bK_{y,t}$ is trace preserving. For each $y$ and density operator $\xi$, $\tr{\bK_{y,t}(\xi)}$ is the probability to measure $y$ at time $t$ knowing  that the quantum state at $t$ is $\xi$. The  quantum filter reads:
\begin{equation}\label{eq:Qfilter}
{\rho}_{t+1}^{(n)} = \frac{\bK_{y_t^{(n)}, t}
 \left({\rho}_{t}^{(n)}\right)}{\tr{\bK_{y_t^{(n)}, t} \left({\rho}_{t}^{(n)}\right)}}
 , ~t=0,\ldots, T_n.
\end{equation}
A discrete-time quantum filter has this  Markovian structure, with $\bK_{y,t}$ depending only on its physical settings.
As the initial state $\hrho$  is unknown, one can only work with $\rho_t^{(n)}$ generated by the above recurrence and starting from a guess $\rho_0^{(n)}=\rho$.

To each measurement record  $(y_t^{(n)})_{t = 0, \ldots, T_n}$, we can associate a number $\mathbb{P}_n(\rho)$ being the probability of getting this record, assuming the initial state was $\rho_0^{(n)}=\rho$. Since $\tr{\bK_{y_t^{(n)}, t} \left(\rho_{t}^{(n)}\right)}$ is the probability  of having detected $y_t^{(n)}$ knowing $\rho_{t}^{(n)}$,  a direct  use of Bayes law yields
$
\mathbb{P}_n(\rho) = \prod_{t=0}^{T_n} \tr{\bK_{y_t^{(n)}, t} \left(\rho_{t}^{(n)}\right)}$  with $\rho_t^{(n)}$ generated thanks to~\eqref{eq:Qfilter}. Some elementary computations show that:
$$
\mathbb{P}_n(\rho) = \tr{\bK_{y_{T_n}^{(n)}, T_n} \circ \ldots \circ \bK_{y_0^{(n)}, 0} \left(\rho\right)}.
$$
Since  the $N$ measurement records are independent realizations from the same initial state,  the probability $\PP{Y~|~\rho}$ of the measurement data:
$$
Y=\big\{y_{t}^{(n)}~|~n\in \{1,\ldots,N\},~t\in\{0,\ldots,T_n\}\big\},$$  knowing  the initial state $\rho$,  reads:
 $$
 \PP{Y~|~\rho} = \prod_{n=1}^{N} \mathbb{P}_n(\rho)
 .
 $$

\subsection{Adjoint-state  derivation of the likelihood function}

The adjoint map $\bK^*_{y,t}$  of $\bK_{y,t}$ is defined by $\tr{A\bK_{y,t}(B)}\equiv \tr{\bK^*_{y,t}(A) B}$ for all Hermitian operators $A$ and $B$. Thus:
$$
\mathbb{P}_n(\rho) = \tr{\rho ~\bK^*_{y_{0}^{(n)}, 0} \circ \ldots \circ \bK^*_{y_{T_n}^{(n)}, T_n}(I)},
$$
\AS{where $I$ is the identity operator}.
Consider the normalized adjoint quantum filter with the  adjoint state  $E_t$ (see, e.g.,\cite{Tsang2009PRL,GammeJM2013PRL}),  with final condition $E_{T_n+1}^{(n)} = I $ and  governed by the following backward recurrence:
\begin{equation}\label{eq:AQfilter}
 E_t^{(n)} = \frac{ \bK_{y_{t}^{(n)}, t}^{*} \left(E_{t+1}^{(n)}\right)}{\tr{\bK_{y_{t}^{(n)}, t}^{*} \left(E_{t+1}^{(n)}\right)}}
 \text{ for }t=T_n, \ldots, 0.
\end{equation}
It defines  a family of Hermitian non-negative  operators $(E_{t}^{(n)})$ of trace one and depending only on the measurement data $Y$.
We have   $\bK^*_{y_{0}^{(n)}, 0} \circ \ldots \circ \bK^*_{y_{T_n}^{(n)}, T_n}(I)=  c_n  E_0^{(n)}$ with $c_n$ depending only on $(y_0^{(n)},\ldots,y_{T_n}^{(n)}$).  Thus $\mathbb{P}_n(\rho)= c_n \tr{\rho E_0^{(n)}}$ and we have:
\begin{equation}\label{eq:MaxLikeEn}
\PP{Y~|~\rho} =  \prod_{n=1}^{N} c_n\tr{\rho E^{(n)}},
\end{equation}
where  $E^{(n)}$ stands for $E_0^{(n)}$.  This formula is a  generalization of~\eqref{eq:MaxLikePOVM} where  $E^{(n)}$ replaces $\pi_{n}$.
\subsection{Quantum-state tomography}
The maximum likelihood  (MaxLike) estimate $\rho_{\ML}$ of the   hidden  initial quantum state $\hrho$ underlying the measurement  data $Y$  is given by maximizing the likelihood function $ \PP{Y~|~\rho}$. It is usual to consider the Log-likelihood function $f(\rho)=\log\left(\PP{Y~|~\rho}\right)$:
\begin{multline}\label{eq:LogLikelihood}
f(\rho)=\sum_{n=1}^{N} \log c_n + \sum_{n=1}^{N} \log\left(\tr{\rho E^{(n)}}\right)
	\\ =C + \sum_{n=1}^{N} \log\left(\tr{\rho E^{(n)}}\right)
\end{multline}
where $C$ is a constant independent from $\rho$. Thus,
\begin{equation}\label{eq:rhoML}
  \rho_{\ML} = \underset{\rho \in \mathcal{D}}{\text{argmax}} f(\rho)
\end{equation}
where $ \mathcal{D}$ is the set of density operators. Assume that the  underlying Hilbert space  is  finite dimensional. Then  $\mathcal{D}$ is a closed convex set, and $f$ is a smooth concave function. Thus this optimization problem can be solved  numerically efficiently (see, e.g., \cite{BoydBook2009}). Moreover, $\rho_{\ML}$ is characterized by the following necessary and sufficient conditions: there exists a nonnegative  scalar $\lambda_{\ML}$ such that:
\begin{equation}\label{eq:CNSML}
[\rho_{\ML},\nabla f_{\ML}]=0 \text{ and }  \lambda_{\ML} P_{\ML} \leq  \nabla f_{\ML}\leq \lambda_{\ML} I
\end{equation}
where $P_{\ML}$ is the orthogonal projector on the range of $\rho_{\ML}$ and $\nabla f_{\ML} $ is the gradient at $\rho_{\ML}$  of the log-likelihood:
\begin{equation}\label{eq:GradLogLikelihood}
  \nabla f_\ML=\sum_{n=1}^{N} \frac{E^{(n)}}{\tr{\rho_{\ML} E^{(n)}}}
  .
\end{equation}
The necessary and sufficient condition~\eqref{eq:CNSML} is just the  translation of the standard optimality criterion for  a convex  optimization problem (see, e.g., \cite{BoydBook2009}):  $\rho_\ML$   maximizes the log likelihood function over the convex set of  density operators, if and only if,  for all density operators $\rho$, $\tr{(\rho-\rho_\ML) \nabla f_\ML } \leq 0$.
When $\rho_\ML$ has full rank, it  belongs to the interior of $\DD$. Then $P_\ML=I$ and  $\nabla f_\ML$ is colinear to $I$.

When $\rho_\ML$ is rank deficient,  it  lies on the boundary of $\DD$. Then $P_\ML <  I$  and~\eqref{eq:CNSML} means that the gradient of the log-likelihood is pointing outward $\DD$ and is  orthogonal to the tangent space at $\rho_\ML$  to  the submanifold of density operators with the same rank as $\rho_\ML$.

When the  likelihood function is concentrated  around its maximum,  $\rho_\ML$ appears to be an  approximation of the Bayesian Mean estimate  $\rho_{\BM}$  whose definition has been recalled in~\eqref{eq:rhoBM}.  It  is proved in appendix that $\rho_{\BM}\approx \rho_{\ML}$ independently of the chosen prior distribution $\mathbb{P}_0$. Thus,  for any Hermitian operator $A$, its Bayesian mean:
$$
 \langle A \rangle_\BM = \frac{\int_{\DD} \tr{\rho A}~\exp(f(\rho)) \mathbb{P}_0(\rho)~d\rho}{\int_{\DD}\exp(f(\rho)) \mathbb{P}_0(\rho)~d\rho}
$$
can be approximated by $\langle A \rangle_\BM\approx \tr{\rho_\ML  A}$. Similarly, we prove in appendix that its Bayesian variance:
$$
\sigma^2_{\BM}(A) = \frac{\int_{\DD} \trr{ (\rho-\rho_{\ML}) A} ~\exp(f(\rho)) \mathbb{P}_0(\rho)~d\rho}{\int_{\DD}\exp(f(\rho)) \mathbb{P}_0(\rho)~d\rho}
,
$$
which captures  the mean uncertainty on the value of $\langle A \rangle_\BM$, due to the fact that the hidden initial state $\hrho$ is unknown, can be approximated by:
\begin{equation}\label{eq:Avariance}
  \sigma^2_{\BM}(A) \approx \sigma^2_{\ML}(A) \equiv \tr{A_{\parallel} ~\bR^{-1}(A_\parallel)}
\end{equation}
where $B_{\parallel}= B - \frac{\tr{BP_{\ML}}}{\tr{P_{\ML}}} P_{\ML} - (I-P_{\ML}) B (I-P_{\ML})$ is the orthogonal projector of any  Hermitian operator $B$ on the tangent space at $\rho_\ML$ to the submanifold of Hermitian operators  with zero trace and with ranks equal  to the rank of $\rho_\ML$. Here above,  the linear super-operator $\bR$  reads for any Hermitian operator $X$,
\begin{multline}\label{eq:FisherInfo}
  \bR(X)= \sum_{n=1}^{N} \frac{\tr{ X E^{(n)}_\parallel}}{\trr{ \rho_{\ML} E^{(n)}}} E^{(n)}_\parallel
  \\
+ \tfrac{1}{2} (\lambda_{\ML}I-\nabla f_{\ML}) X \rho_{\ML}^{+} +   \tfrac{1}{2}\rho_{\ML}^{+} X (\lambda_{\ML}I-\nabla f_{\ML})
\end{multline}
with  $\lambda_{\ML}$ and $P_{\ML}$ appearing in~\eqref{eq:CNSML} and $\rho_{\ML}^{+}$ the Moore-Penrose pseudo-inverse of $\rho_{\ML}$.
Notice that the super-operator $\bR$ is symmetric and non-negative for the Frobenius product.  Thus, $\sigma^2_\ML(A) $ is always non-negative as it should be.

When $\rho_\ML$ has full rank, $P_\ML=I$ and  $\nabla f_\ML=\lambda_\ML I$. Then
$\bR(X)$ corresponds to the orthogonal projection (for the Frobenius product) of  $ - \nabla^2 f_\ML(X)$ onto the subspace of Hermitian operators of zero trace. Here, $\nabla^{2}f_\ML$ corresponds to the Hessian of $f$ at $\rho_\ML$:
$$
 \nabla^{2}f_\ML(X)= -\sum_{n=1}^{N} \frac{\tr{ X E^{(n)}}}{\trr{ \rho_{\ML} E^{(n)}}} E^{(n)}
.
$$
We recover, up to this orthogonal projection, the standard Cram\'er-Rao bounds attached to MaxLike estimation:  $-\nabla^{2}f_\ML$ stands for the Fisher information matrix.  The super-operator $\bR$ defined by~\eqref{eq:FisherInfo} is the prolongation  of such   Fisher-information matrix when $\rho_\ML$ lies  on the boundary of $\DD$. Notice its dependence on boundary  curvature   due to the fact that $(\lambda_\ML I -\nabla f_\ML) X \rho_\ML^{+}$ does not vanish in general. Notice that the expressions of $\langle A \rangle_\BM$ and $\sigma^2_{\BM}(A)$ do not depend on the prior distribution of probability $\mathbb{P}_0(\rho)$. This can be easily understood, as these expressions are the first terms of asymptotic expansions when $f(\rho)$ grows large, i.e. when the amount of information brought by the measurement records make the initial information $\mathbb{P}_0(\rho)$ outdated.  The computations underlying~\eqref{eq:FisherInfo} are given in appendix: they rely on a specific  application of asymptotic expansions for multi-dimensional Laplace integrals given in~\cite[chapter 8]{BleistenHandelsmanBook}.

\subsection{Experimental validation for QND photon counting}

We apply in the following the MaxLike reconstruction to the state $\hrho$ of the light field stored in a cavity based on the experiment reported in~\cite{PastQuantumStateLKB2014} and, more precisely, experimental data associated to Fig. 4b therein. The field of a very high-quality superconducting cavity (frequency of 51~GHz, photon lifetime of $T_c=65$~ms) is initially in a thermal state (temperature of 0.8~K, mean number of thermal photons of $n_b=0.06$). At time $t=0^{-}$, a single photon is injected into the field. Due to experimental imperfections, in reality $1.26$ photons are injected on average. The QND photon number measurement consists of a long sequence of atomic probes (samples of individual atoms prepared in a highly excited circular Rydberg state) crossing the cavity mode one by one and separated by 86~$\mu$s. The measurement starts at $t  = -172$~ms and has a duration of $344$~ms, large compared to the photon lifetime.  The main sources of measurement imperfections are random atomic occupation of samples, non-constant atom-photon interaction from sample to sample, non-ideal atom state detection, etc. The decoherence is mainly due to the limited cavity lifetime $T_c$ leading to photon losses. For more details on the considered experiment, please refer to~\cite{PastQuantumStateLKB2014} and references therein.

Here, the quantum and adjoint states $\rho_t$ and $E_t$ are diagonal in the Fock basis and truncated to a maximum of $7$ photons: they are described by vectors of dimension 8. The partial Kraus maps $\bK_{y,t}$ reduced to $8\times 8$ real matrices and the computations of the $E^{(n)}$ rely on the transpose of these real matrices. We do not detail here the precise expressions of the different $\bK_{y,t}$: they can be deduced from~\cite{PastQuantumStateLKB2014}. For any $t$ between $-100$~ms and $+150$~ms, we compute the MaxLike estimation $\rho_\ML(t)$ of $\hrho(t)$, the state at time $t$, based on the measurement outcomes between $t$ and $+172$~ms.
\AS{In a tomographic spirit, this MaxLike reconstruction takes into account the precise model of the ``measurement'' of state $\hrho(t)$: QND interaction with the atomic probes while the cavity is decohering, between time $t$ and our last information at $+172$~ms. However, it does not assume anything about the target state prior to measurement, i.e.~it neglects anything that happened before $t$. Moreover, the addition of a photon at $t=0^-$ is not taken into account in the model, which means that for all $t<0$ there is in fact a mismatch in the model of the measurement process lasting from $t$ to $+172$~ms; this allows us to get an idea about the robustness of $\rho_\ML$ to model errors.} For any time $t$, the rank of $\rho_\ML(t)$ is strictly less than $8$ (between $2$ and $5$) and we have checked that it satisfies the characterization~\eqref{eq:CNSML}.

\begin{figure}[htb]
  {\includegraphics[width=\columnwidth]{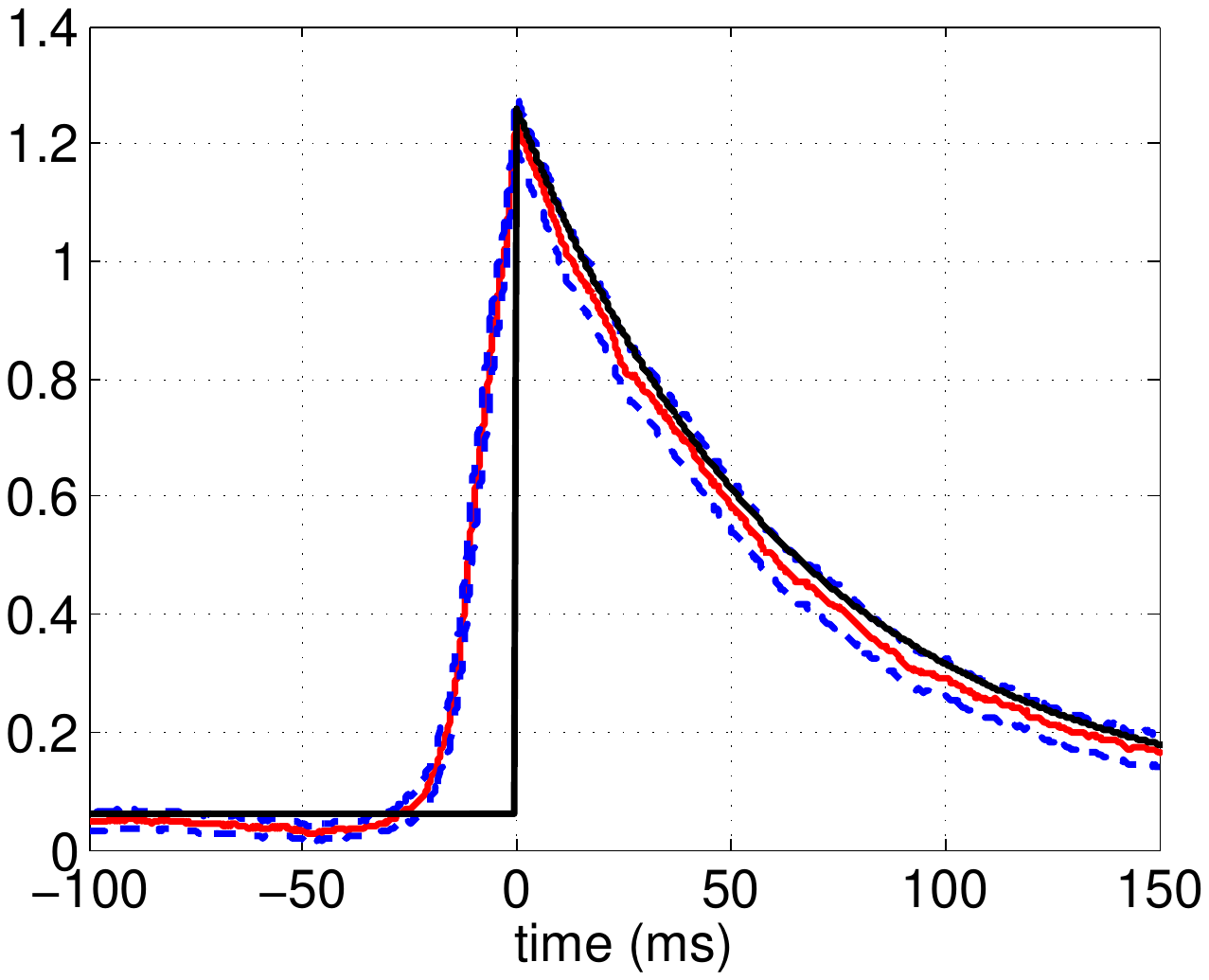}}
  {\includegraphics[width=\columnwidth]{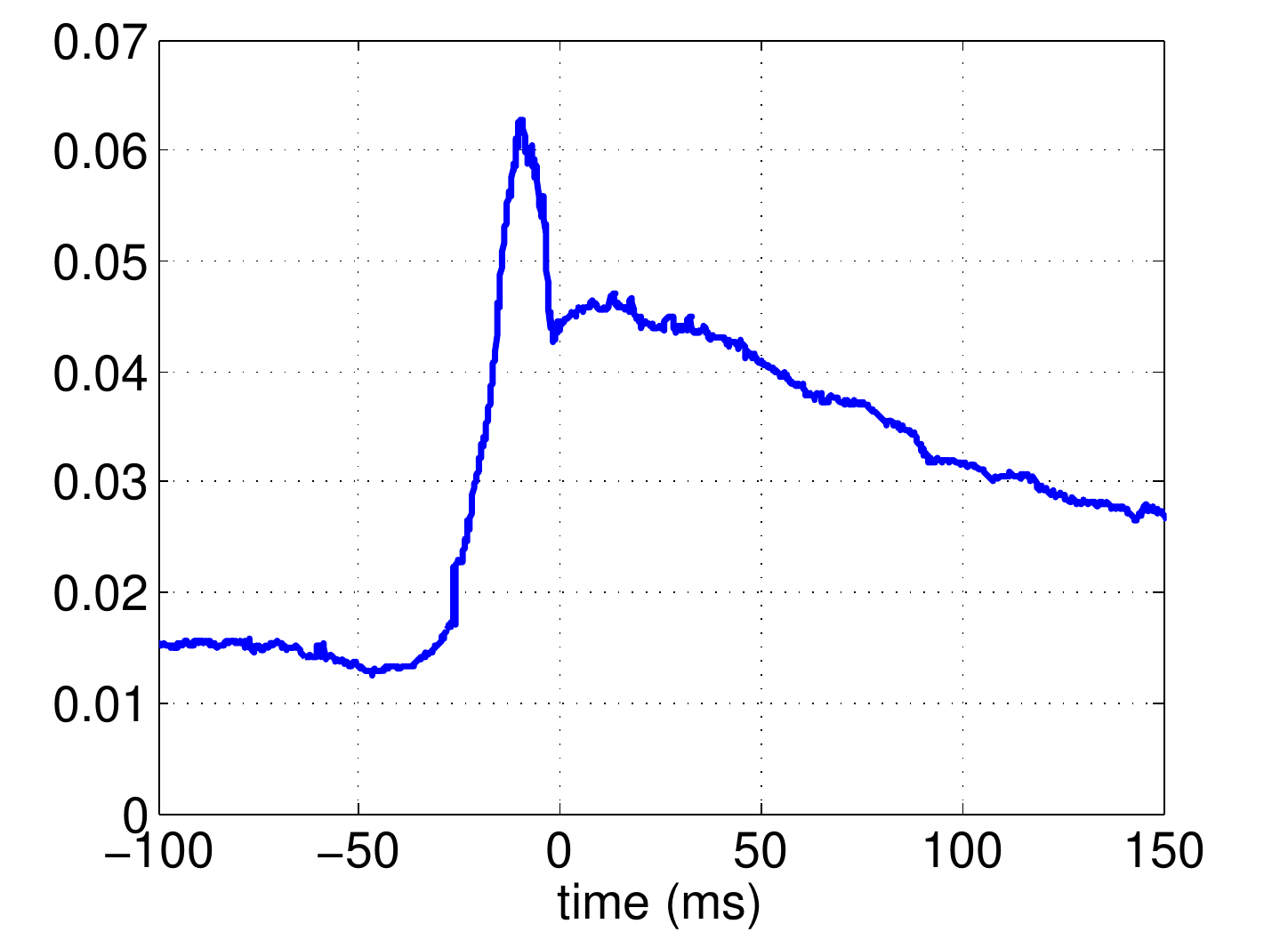}}
  \caption{(Color online)  MaxLike quantum-state tomography based on the QND photon
		measurements reported in~\cite[figure~4b]{PastQuantumStateLKB2014} for a
		photon creation quantum jump induced at $t=0$. The latter is not taken into account in the measurement model. Top: solid red line corresponds to the MaxLike mean photon-number $\tr{\rho_\ML(t) \ba^\dag \ba }$ \AS{of the tomography target $\hrho(t)$, and} based on measurements between $t$ and $+172$~ms. The two dashed blue lines correspond to $\tr{\rho_\ML \ba^\dag \ba}\pm 2\sigma_\ML(\ba^\dag \ba)$, a $95\%$ confidence interval. The black line corresponds to the mean photon-number $\bket{\ba^\dag \ba }_t=\tr{\hrho(t)\ba^\dag\ba}$ given by~\eqref{eq:Nmean}. Bottom: $2\sigma_\ML(\ba^\dag \ba)$  based on~\eqref{eq:Avariance}. See text for more explanations.
    }
     \label{fig:TomoOnePhoton}
\end{figure}

The results for $N=1390$ measurement trajectories are illustrated on figure~\ref{fig:TomoOnePhoton}.  In order to evaluate  the precision of such  MaxLike estimation, we have computed  the variance given in~\eqref{eq:Avariance} for the photon-number operator $\ba^\dag \ba$. The resulting  $95\%$ confidence intervals  show that,  for all $t$ between $-100$ and $+150$~ms, the mean photon numbers  are evaluated with estimated  uncertainties  between $\pm 0.015$ and $\pm 0.065$. \AS{We also compare the MaxLike reconstruction for $t$ between $-100$ and $+150$~ms with the mean photon-number
$\bket{\ba^\dag\ba}_t=\tr{\hrho(t) \ba^\dag \ba }$ given by:
\begin{equation}\label{eq:Nmean}
\bket{\ba^\dag \ba }_t= \left\{
                          \begin{array}{ll}
                            n_b, & \hbox{for $t < 0$;} \\
                           n_b+ (\bket{\ba^\dag \ba }_0-n_b) e^{-t/T_c}, & \hbox{for $t\geq 0$.}
                          \end{array}
                        \right.
\end{equation}
where $\bket{\ba^\dag \ba }_0=1.26$, $T_c=65$~ms and $n_b=0.06$ are derived from~\cite{PastQuantumStateLKB2014}.}
We observe that $\bket{\ba^\dag \ba }_t$ remains almost inside the $95\%$-confidence tube except  for $t$ between $-50$~ms and $0$. \AS{This is a very positive result and incidentally, much better than the ``backward reconstruction'' based on future measurements performed in \cite[Fig.4(b), blue curve]{PastQuantumStateLKB2014}, which features a significant offset. Let us briefly comment on the MaxLike estimation at each $t$.}

At time $0$~ms, the MaxLike photon number reaches its maximum, $1.237\pm 0.045 $. It corresponds to the programmed injection of~\cite{PastQuantumStateLKB2014}. \AS{Our MaxLike estimation is a tomography of that state, i.e.~it never includes information about what we have injected into the system before $t=0$, and hence this is an independent confirmation of the programmed injection. For $t>0$, we expect the mean photon number to decay due to the finite photon lifetime. The MaxLike estimate confirms such behavior --- again, just by making the tomography of $\hrho(t)$ using measurements obtained after $t$ i.e.~without any information on how $\hrho(t)$ might have been constructed. For $t$ between $-50$~ms and $0$, the observed mismatch is expected, since the photon jump at $t=0$ is not taken into account in the tomography model for the MaxLike reconstruction. Yet for $t<-50$~ms we get a MaxLike mean photon number of $0.042\pm 0.015$, close to the thermal photon number $0.06$ observed in~\cite{PastQuantumStateLKB2014}. This is a notable result, illustrating that the MaxLike estimate is robust to the erroneous tomography model which neglects the photon injection at $t=0$. In other words, our MaxLike estimate appropriately gives more credit to measurements obtained right after $\hrho(t)$ was prepared, than to measurements further in the future; although it takes all these measurements into account with a (supposedly) appropriate weight. Hence, when $t$ becomes negative, one simply has to collect enough measurements between $t$ and $0$ to recover a correct MaxLike estimation of $\hrho(t)$. Due to the moderate amount $N$ of measurement trajectories here, having sufficiently many measurements means requiring that $t<-50$~ms.}

\section{ Adaptation to continuous-time} \label{sec:continuous}

 Most open quantum systems are modeled in continuous-time. This section shows how to recover and exploit  the  discrete-time setting  when  $dt$, the  sampling time,  is  much smaller than the characteristic  time-scales for  systems modeled  with  stochastic master differential equations driven by Wiener processes.
\subsection{Diffusive quantum filtering and adjoint-state }

The stochastic master equations admit here  the following form (see, e.g.,~\cite{Belavkin1992,BarchielliGregorattiBook}):
{\small\begin{multline}\label{eq:SME}
d\rho_t=\left(-\tfrac{i}{\hbar}[H,\rho_t]+\sum_\nu  L_\nu \rho_t  L_\nu^\dag - \tfrac{1}{2}( L_\nu^\dag  L_\nu\rho_t+\rho_t  L_\nu^\dag  L_\nu)\right) dt
\\
+ \sum_\nu \sqrt{\eta_\nu}\left( L_\nu\rho_t+\rho_t  L_\nu^\dag-\tr{( L_\nu+ L_\nu^\dag)\rho_t}\rho_t \right) dW_{\nu,t}
\end{multline}}
with    $dW_{\nu,t}= dy_{\nu,t}- \sqrt{\eta_\nu}\tr{( L_\nu+ L_\nu^\dag)\,\rho_t}\,dt $ given by the measurement $y_{\nu,t}$,  attached to  the  operator $L_\nu$,  with efficiency $\eta_\nu$ between $0$ and $1$.  Here, $dW_{\nu,t}$ are independent scalar Wiener processes and $H$ is the Hamiltonian  that could depend on time $t$ via, e. g., some  time-varying  coherent drives.

Following~\cite{RouchR2015PRA,CampagnePhD2015,SixCDC2015}, such quantum filters admit the following  infinitesimal discrete-time formulations based on  It$\bar{\text{o}}$ rules:
$$
\rho_{t+dt}= \frac{\bK_{dy_t}(\rho_t)}{\tr{  \bK_{dy_t}(\rho_t)}},
$$
where the  complete positive maps   $ \bK_{dy_t}$  depend on  $dy_t=(dy_{\nu,t})$    according to:
$$
\bK_{dy_t}(\rho_t)=  M_{dy_t} \rho_t  M_{dy_t}^\dag + \sum_\nu (1-\eta_\nu )  L_{\nu}\rho_t  L_{\nu}^\dag dt
$$
with
{\small $$ M_{ dy_t} =  I + \left(-\tfrac{i}{\hbar} H -  \tfrac{1}{2} \left(    \sum_\nu L_\nu^\dag   L_\nu \right)\right) dt +
\sum_\nu \sqrt{\eta_\nu}  dy_{\nu,t}   L_\nu.
$$}
The  probability  of outcome  $ dy_t=(dy_{\nu,t})$ is given then by the following distribution based on Gaussian laws of variance $dt$:
{\small \begin{multline*}
  \PP{  dy_t  \in\prod_{\nu} [\xi_\nu,\xi_\nu+d\xi_\nu]~\Big|~\rho_t}
  \\ =\tr{  \bK_{ \xi}(\rho_t)}~\prod_\nu  e^{- \xi_\nu^2/2dt}\frac{d\xi_\nu}{\sqrt{2\pi dt}}
  .
\end{multline*}}

Take a sampling-time $dt$ much smaller than the time-constant involved in the Hamiltonian  $H$ and in the decoherence operator $L_\nu$. Then we can exploit the above  formulations similar to  discrete-time quantum filtering and compute  the normalized adjoint states $E_t^{(n)}$. They are  associated to the measurement data $(y_{t}^{(n)})_{0\leq t \leq T_n}$ corresponding  to the quantum trajectory  number $n$, via the following discrete-time formulation:
$$
E_{t}=\frac{ \bK_{dy_{t}^{(n)}}^{*} \left(E_{t+dt}\right)}{\tr{\bK_{dy_{t}^{(n)}}^{*} \left(E_{t+dt}\right)}}
\text{ with } E_{T_n+dt}=I/\tr{I}
$$
with $T_n=\mathcal{N}_n dt$ for $\mathcal{N}_n$ a large integer  and where  the adjoint $\bK^*_{dy_t}$ of $\bK_{dy_t}$ reads:
$$
\bK^*_{dy_t}(\rho_t)=  M_{dy_t}^\dag \rho_t  M_{dy_t}+ \sum_\nu (1-\eta_\nu )  L_{\nu}^\dag \rho_t  L_{\nu}dt
.
$$
 After having obtained, for the $N$  quantum trajectories, the value at $t=0$ of the adjoint states $(E^{(n)})_{n=1,\ldots,N}$, we  can directly use the  MaxLike quantum-state tomography developed in previous section.

\subsection{Quantum-state tomography for a qubit}

For a two-level system,  the MaxLike estimation developed in the previous sections admits a simpler formulation with the Bloch sphere variables that can be used  for both  $\rho$ and $E$:
\begin{align*}
\rho&=\frac{I + x \sigma_x  + y \sigma_y + z \sigma_z}{2},
\\
E&=\frac{I + e_x \sigma_x  + e_y \sigma_y + e_z  \sigma_z}{2},
\end{align*}
where   $(x,y,z)$ and $(e_x,e_y,e_z)$ correspond to the coordinate of vectors with $\sigma_x$, $\sigma_y$ and $\sigma_z$  the three Pauli matrices.  Here   the convex set $\DD$ corresponds to the unit ball  $x^2+y^2+z^2 \leq 1$ and the Frobenius product between operators to the  Euclidian product between vectors  in  the 3-dimensional Euclidian space.  Then, the gradient of the log-likelihood function~\eqref{eq:GradLogLikelihood} becomes the vector:
$$
 \nabla f_\ML = \sum_{n} \tfrac{1}{1+x_\ML e_x^{(n)} + y_\ML e_y^{(n)}  + z_\ML e_z^{(n)}  }
  \begin{pmatrix} e_x^{(n)} \\e_y^{(n)} \\e_z^{(n)} \end{pmatrix}
  .
$$
The characterization of $\rho_\ML$ given in~\eqref{eq:CNSML} becomes as follows: if $x_\ML^2+y_\ML^2 + z_\ML^2 < 1$, then $\nabla f_\ML=0$;
if $x_\ML^2+y_\ML^2 + z_\ML^2 =1$ then $\nabla f_\ML= \lambda_\ML \begin{pmatrix} x_\ML \\ y_\ML \\ z_\ML \end{pmatrix}$ with $\lambda_\ML\geq 0$.
The super-operator $\bR$  defined in~\eqref{eq:FisherInfo}   becomes  a $3\times 3$ symmetric non-negative  matrix. It reads as follows:
\begin{itemize}
  \item when $ x_\ML^2+y_\ML^2 + z_\ML^2 < 1$,  we have:
  {\small
  $$
  \bR=\sum_n \frac{
  \begin{pmatrix}
    e_x^{(n)}e_x^{(n)}  & e_x^{(n)} e_y^{(n)} & e_x^{(n)} e_z^{(n)} \\
    e_y^{(n)} e_x^{(n)}  & e_y^{(n)} e_y^{(n)} & e_y^{(n)} e_z^{(n)} \\
    e_z^{(n)}e_x^{(n)}  & e_z^{(n)}e_y^{(n)} & e_z^{(n)}e_z^{(n)}
  \end{pmatrix}
  }{\left(1+x_\ML e_x^{(n)} + y_\ML e_y^{(n)}  + z_\ML e_z^{(n)}\right)^2 }
  $$}
  It is usually of rank $3$ and  can be inverted on any operator of the form   $A=\frac{a \sigma_x + b \sigma_y + c\sigma_z}{2}$ associated to the vector $ \begin{pmatrix} a\\b\\c \end{pmatrix}$ to get a variance  estimation  via~\eqref{eq:Avariance} where the trace is replaced by the Euclidean scalar product.

  \item When $ x_\ML^2+y_\ML^2 + z_\ML^2 = 1$ and $\lambda_\ML >0$ large enough, we have:
     {\scriptsize   \begin{multline*}
     \bR=\sum_n \frac{
  \begin{pmatrix}
    e_{\parallel x}^{(n)}e_{\parallel x}^{(n)}  & e_{\parallel x}^{(n)} e_{\parallel y}^{(n)} & e_{\parallel x}^{(n)} e_{\parallel z}^{(n)} \\
    e_{\parallel y}^{(n)} e_{\parallel x}^{(n)}  & e_{\parallel y}^{(n)} e_{\parallel y}^{(n)} & e_{\parallel y}^{(n)} e_{\parallel z}^{(n)} \\
    e_{\parallel z}^{(n)}e_{\parallel x}^{(n)}  & e_{\parallel z}^{(n)}e_{\parallel y}^{(n)} & e_{\parallel z}^{(n)}e_{\parallel z}^{(n)}
  \end{pmatrix}
  }{\left(1+x_\ML e_x^{(n)} + y_\ML e_y^{(n)}  + z_\ML e_z^{(n)}\right)^2 }
  \\
+
  \lambda_\ML
  \begin{pmatrix}
   1- x_\ML x_\ML  & -x_\ML y_\ML & -x_\ML z_\ML \\
   - y_\ML  x_\ML  & 1-y_\ML y_\ML & -y_\ML z_\ML  \\
    -z_\ML x_\ML   & -z_\ML y_\ML & 1-z_\ML z_\ML
  \end{pmatrix}
      \end{multline*}}
  where  $e_{\parallel \xi}^{(n)} = e_\xi^{(n)}- s^{(n)} \xi_\ML$ for $\xi=x,y,z$ and $s^{(n)}=e_x^{(n)}x_\ML+e_y^{(n)}y_\ML+e_z^{(n)}z_\ML$.
    Its rank is less than or equal to $2$. The inverse of $\bR$ appearing in~\eqref{eq:Avariance} corresponds here to the Moore-Penrose pseudo-inverse. It is evaluated on  the vector associated to $A_\parallel$:
      $$
      A_\parallel = \tfrac{(a- s x_\ML )\sigma_x+ (b-s y_\ML)\sigma_y + (c-s z_\ML)\sigma_z}{2}
      $$
      with $s=a x_\ML + b y_\ML + c z_\ML$.

\end{itemize}

\subsection{Experimental  validation for a qubit with fluorescence heterodyne measurements}

MaxLike quantum-state tomography is conducted on a superconducting qubit  whose fluorescence field is measured using a heterodyne detector~\cite{CampagneEtAlPRX2015}. For the detailed  physics of this  experiment, see~\cite{CampagneEtAlPRL2014,CampagnePhD2015}. \AS{The measurement model is described by} a stochastic master equation of the form~\eqref{eq:SME} with $H=0$ and $\nu=1,2,3$:
$$
L_1= \sqrt{\tfrac{1}{2T_1}} \frac{\sigma_x-i\sigma_y}{2}, \quad  L_2=i L_1, \quad L_3=\sqrt{\tfrac{1}{2T_\phi}} \sigma_z
$$
with $\eta_1=\eta_2=0.24 $ the  efficiency of the heterodyne measurement and with $\eta_3=0$ corresponding  to an unmonitored dephasing channel. The measurement and dephasing time constants  are $T_1=4.15~\mu$s and $T_\phi=35~\mu$s.

We have at our disposal $N=4.10^{4}$ quantum trajectories with the same length  $T_n=T=9.2~\mu$s with a sampling-time  $dt= 200~\text{ns}\approx \frac{1}{20}T_1$. \AS{Each trajectory is supposed to} start at time $t=0$ from the same initial state $\hrho_0$  close to $(\ket g + \ket e)/\sqrt{2}$. \AS{In a first test, probably closest to experimental needs, the goal is to check this fact by performing a MaxLike tomographic estimation of $\hrho_0$ based on records of the continuous fluorescence signals between $0$ and final time $T$}.
For the  trajectory number $n$, the measurement record corresponds to  $2\times 47$  real values corresponding to $\int_{t-dt}^t dy^{(n)}_{1}$ and $\int_{t-dt}^t dy^{(n)}_{2}$ for $t=dt,2dt,\ldots, 47 dt$.  From the measurements between $0$ and $T$, we get an estimation of the quantum state $\hrho_0$ with a  $95\%$ confidence interval using the above formula for  $\sigma^2_\ML(A)$ and $A=\sigma_x$, $\sigma_y$ and $\sigma_z$:
\begin{align*}
  \tr{\hrho_0 \sigma_x} &= 0.99 \pm  0.06 \\
  \tr{\hrho_0 \sigma_y} &=-0.03\pm 0.07 \\
  \tr{\hrho_0 \sigma_z} &=  -0.10 \pm 0.19
\end{align*}
This estimated value of the initial state is consistent with a gate error of a few percent in the preparation of $(\ket g + \ket e)/\sqrt{2}$ starting from the thermal state with less than 1\% excitation.

{To further validate how the MaxLike tomography can reconstruct different states of the qubit we next perform, as for the previous experiment, the tomography of the state $\hrho_t$ obtained at various times $t$, using $N$ records of the continuous fluorescence signals between time $t$ and final time $T$. Figure~\ref{fig:TomoQubitXYb} shows,  for $t$ between $0$ and $5~\mu$s, the resulting estimates $x_\ML(t)$ and $y_\ML(t)$ (red solid lines) with their  $95\%$ confidence intervals (blue dashed  lines). The black circle marks correspond to average over a much larger set of $N^*=3.10^{6}$  trajectories  of the normalized signals,  $ \sqrt{\tfrac{2 T_1}{\eta}}\int_{t-dt}^tdy_{1}$ and $\sqrt{\tfrac{2 T_1}{\eta}} \int_{t-dt}^{t} dy_{2}$. \AS{Since $ \langle dy_{\nu,t} \rangle = \sqrt{\eta_\nu}\tr{( L_\nu+ L_\nu^\dag)\rho_t}dt$, these black circles are meant to provide reliable estimations of  $\tr{\hrho_t \sigma_x}$ and $\tr{\hrho_t\sigma_y}$ at times $t=0,dt,2dt,\ldots, 25 dt$, where $\hrho_t$ starts at $\hrho_0$ and follows the Lindblad master equation corresponding to unread measurements. Thus the message is that our MaxLike method appears to obtain consistent reconstructions, as its confidence interval covers the very precise statistics (black circles) of a more standard model.}

Figure~\ref{fig:TomoQubitZ} corresponds, for the same set of $N$ measurement trajectories,  to the  MaxLike reconstruction  of $z$. Contrarily to $x$ and $y$, this Bloch coordinate cannot be recovered directly by an ensemble average \AS{of the measurement signal, so there are no black circles for validation.}

On Figure~\ref{fig:TomoQubitXYc} we use the same set of $N=4.10^{4}$ measurement records to compute the ensemble average of the normalized heterodyne signals (black circles). \AS{With this smaller dataset, the black circles do not provide a statistically accurate estimation of  $\tr{\hrho_t \sigma_x}$ and $\tr{\hrho_t\sigma_y}$ anymore. To the contrary, the noise attached to such ensemble average is large compared to MaxLike estimation of $x$ and $y$. This shows that as expected, our MaxLike estimation by exploiting all the data between $t$ and $T$ has superior statistical power, compared to an estimation of $\hrho_t$ from measurement outputs obtained at time $t$ only}.

\begin{figure}
  \includegraphics[width=\columnwidth]{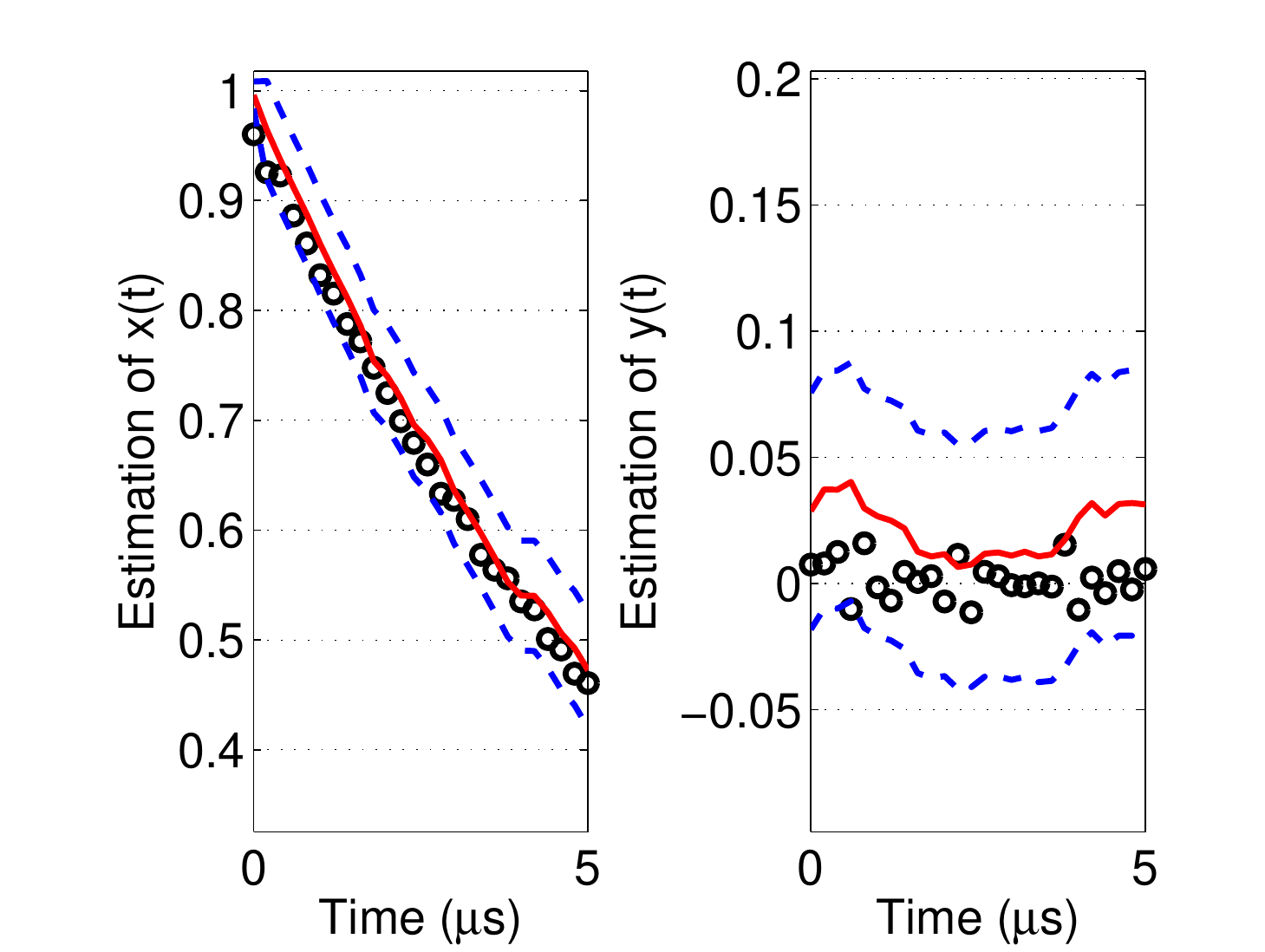}\\
  \caption{(Color online) Comparison of   MaxLike reconstruction  for $x$ (left plot: solid and dashed curves) and $y$ (right plot: solid and dashed curves)  exploiting the measurement data set associated to  $N=4.10^{4}$ experimental  trajectories; with an  ensemble average of the normalized fluorescence  signals (black circle)  exploiting  a much larger data  set associated to $N^*=3.10^{6}$ experimental trajectories; the red solid lines correspond to $x_\ML$ and $y_\ML$; the blue dashed lines   to the confidence intervals $x_\ML \pm 2 \sigma_\ML(\sigma_x)$ and $y_\ML \pm 2 \sigma_\ML(\sigma_y)$ obtained from~\eqref{eq:Avariance}. \AS{The black circles are generally admitted to represent reliable estimates of $\tr{\hrho_t \sigma_x}$ and $\tr{\hrho_t\sigma_y}$; see text for more explanations.}}
    \label{fig:TomoQubitXYb}
\end{figure}
\begin{figure}
  \includegraphics[width=\columnwidth]{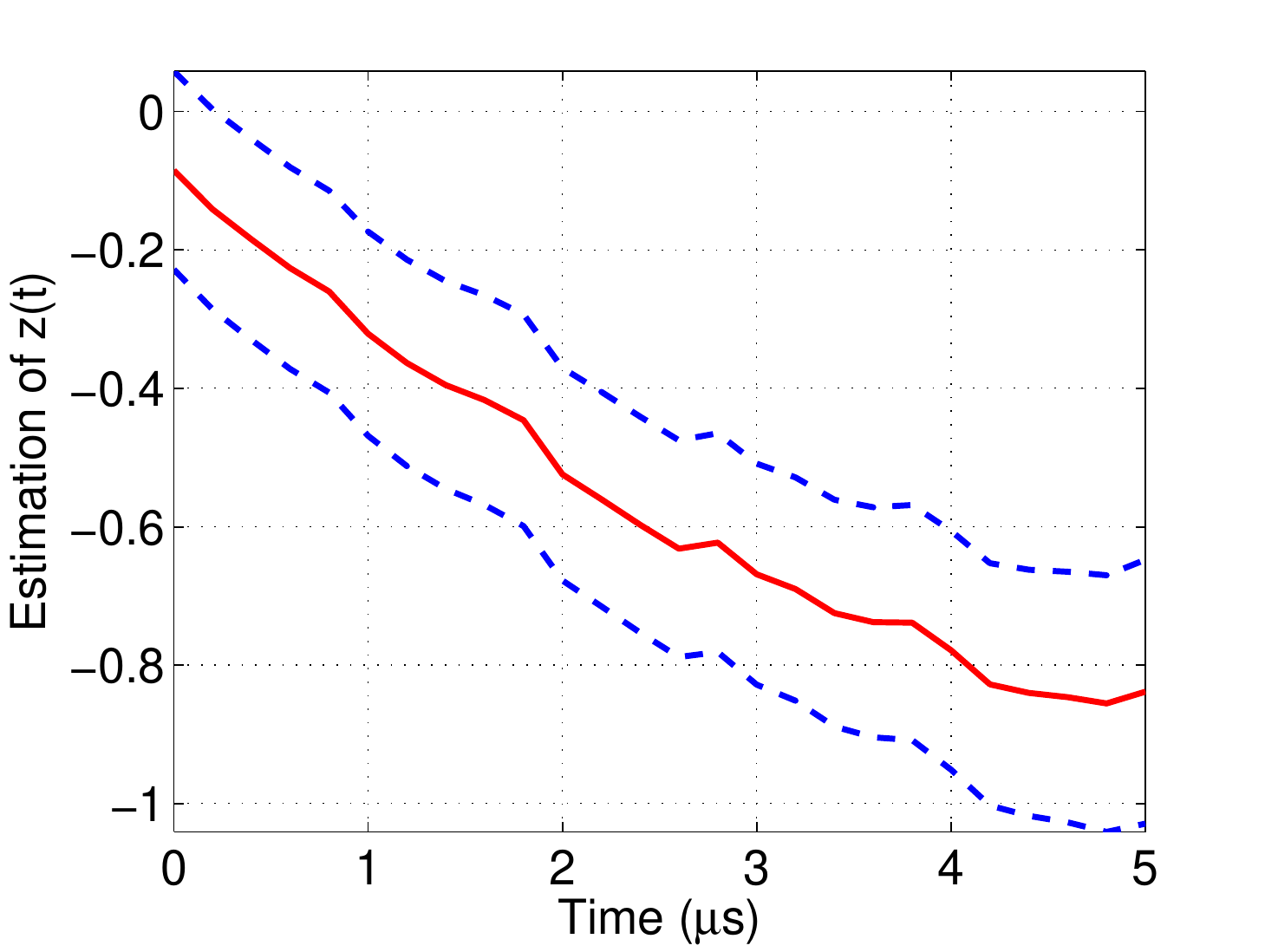}\\
  \caption{(Color online) MaxLike reconstruction  for $z$ exploiting a measurement data set  with $N=4.10^{4}$ experimental  trajectories; the red solid line corresponds to $z_\ML$ and the  blue dashed one   to the confidence interval $z_\ML \pm 2 \sigma_\ML(\sigma_z)$ obtained from~\eqref{eq:Avariance}; see text for more explanations.}
    \label{fig:TomoQubitZ}
\end{figure}
\begin{figure}
  \includegraphics[width=\columnwidth]{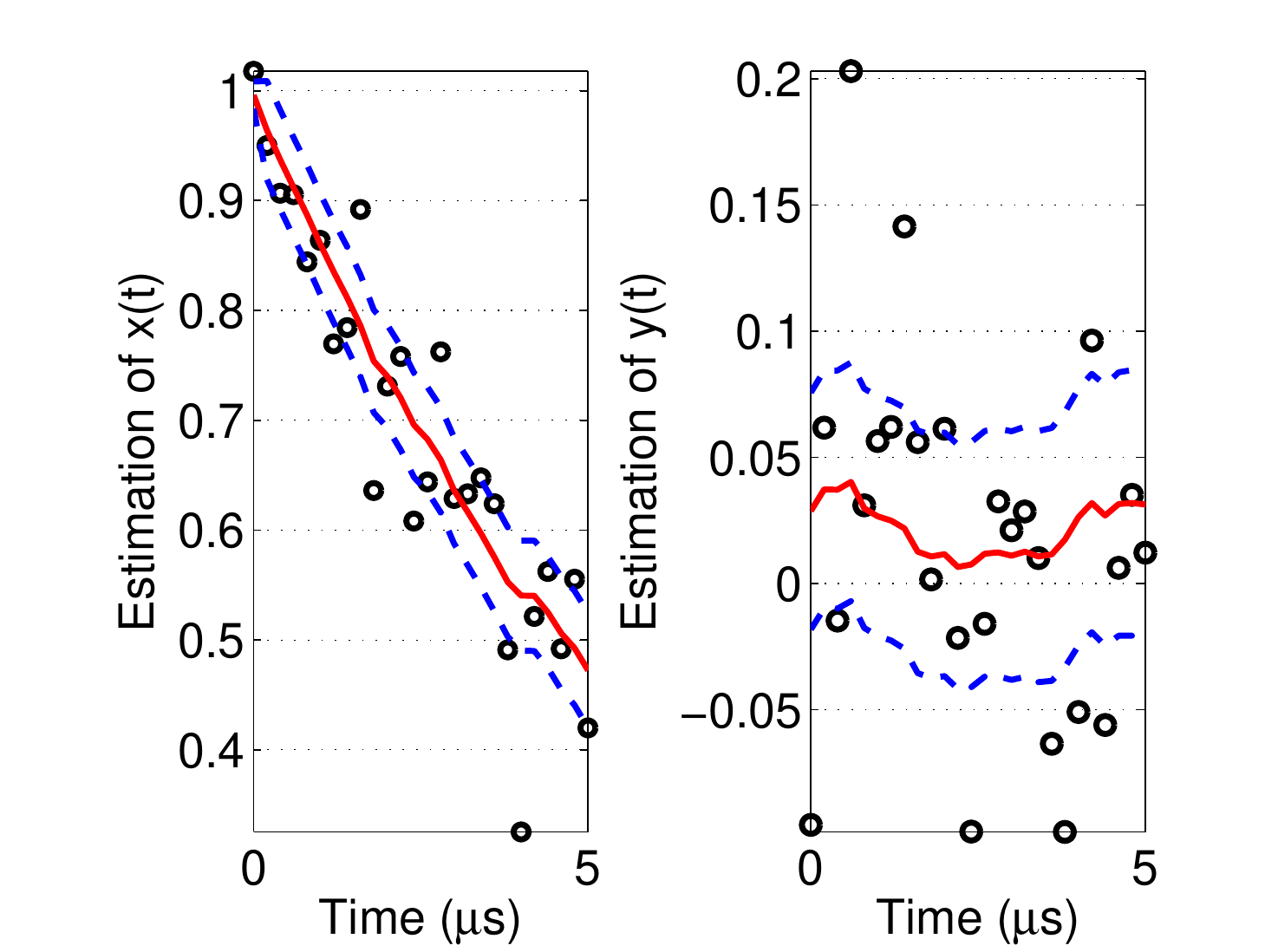}\\
  \caption{(Color online) Similar to figure~\ref{fig:TomoQubitXYb} but the ensemble average of the normalized fluorescence  signals (black circles) use the same dataset as the MaxLike reconstruction. The MaxLike reconstruction appears to be significantly less noisy, illustrating the efficiency of our method; see text for more explanations.}
    \label{fig:TomoQubitXYc}
\end{figure}

\section{Concluding remarks}

We have proposed a new method to take into account imperfection and decoherence for quantum-state tomography based on non-instantaneous measurements.  This method is well adapted to MaxLike estimation since it provides directly the likelihood function. \AS{The latter is written in terms of a sort of back-propagated POVM operators which depend on the measurement trajectory, and which characterize the information the trajectory of outputs carries on about the initial state to be estimated.} We have given an approximation of the confidence interval to complete the MaxLike state-estimate.
\AS{In essence, given many copies of an initial quantum state, we can run it through any well-characterized device giving a sequence of outputs for each copy, and, from this we can estimate efficiently what the initial state was, with an associated confidence interval. The estimator appears to mostly rely on measurement outputs obtained just after the state was prepared, much like a standard POVM; however, it can compensate for "missing statistical power" by relying on further evolution of the trajectory. This seems to guarantee robustness to uncertainties in the model of the dynamic measurement apparatus, as illustrated in our experimental validation.}

The proposed method is directly applicable to \AS{reconstruct the state $\rho_0$ from a standard projective measurement performed on $\rho_t$ with $t>0$. If on the measurement run $n$ the final projective measurement at $T_n$ yields the known state $\rho_n$, then the backward computations~\eqref{eq:AQfilter} just start with $E_{T_n}^{(n)}= \rho_n$ instead of $E_{T_n}^{(n)} = I/\tr{I}$. If measurements are obtained between as the state evolves from $\rho_0$ to $\rho_t$, one just applies our standard filter with this different initial condition. If no measurements are obtained between $0$ and $t$, e.g.~the quantum state just undergoes decoherence, then one can obtain the corresponding filter by modeling the decoherence as hypothetical unread measurements in the environment, and replacing the corresponding $\bK^*_{y_t,t}$ associated to read measurements by the unital linear map $\bK^*_t=\sum_{y} \bK^*_{y,t}$. In this case, the reconstructed MaxLike estimation will appropriately take into account that the POVM was performed not on $\rho_0$, but on a state $\rho_t$ that has evolved from $\rho_0$ according to some known dynamics.}

Extension  of the proposed method to  measurement protocols where a meter is coupled to the system of interest and, after a small time a (projective) measurement on the  meter only is done, appears to be possible. For  example, it could be useful for the Wigner tomography of a quantum oscillator, such as those realized, e.g., in~\cite{LeghtTPKVPSNSHRFSMD2015S}, to take into account higher order Hamiltonian distortions and some decoherence effects during the joint evolution of system and meter.


\appendix

\section{Asymptotic expansions underlying~\eqref{eq:Avariance}}
To formalize the fact that  the likelihood function $\rho \mapsto e^{f(\rho)}$ is concentrated around its maximum, we set $f(\rho)= \Omega \hf(\rho)$, $\Omega >0$ and large with  $\hf(\rho)$  a  normalized log-likelihood with bounded variations:  for any density operators $\rho_1$ and $\rho_2$, $|\hf(\rho_1)-\hf(\rho_2)|\leq 1$.
Throughout this appendix, we assume that $\hf(\rho)$ is maximal for a unique density matrix  $\rho_\ML$ and that its Hessian matrix
$\nabla^2 \hf_\ML$ is negative definite. The dimension of the underlying finite-dimensional  Hilbert space is denoted  here by the integer~$n$.

For any  scalar function $g$ of $\rho$, we consider thus the asymptotic expansion versus $\Omega$ of the following multi-dimensional integral of Laplace type:
$$
I_g(\Omega) = \int_{\DD} g(\rho) e^{\Omega \hf(\rho)} d\rho
.
$$
The domain of integration $\DD$ is a compact convex subset of  the set  of $n \times n$  Hermitian matrices  of  trace one. Here, $d\rho$ stands for the standard Euclidian volume element on $\DD$,  derived from the Frobenius product between $n\times n$ Hermitian matrices. We denote by  $\DD_{n-1}$  the set  of density matrices  of rank less or equal  to  $n-1$. It corresponds to the boundary of $\DD$. Here, we denote by $\nabla \hf$ and $\nabla^2 \hf$ the gradient and Hessian of $\hf$ that is considered  as a scalar  function on the set of Hermitian matrices of unit-trace.

Assume that $\rho_\ML$ has full rank, i.e.,  $\hf$ reaches its maximum in the interior of $\DD$ and $\nabla \hf_\ML=0$. Then we can use~\cite[eq. 8.3.52]{BleistenHandelsmanBook} to get the following equivalent for $I_g(\Omega)$ using that $\dim \DD=n^2-1$:
\begin{equation}\label{eq:Ig0}
  I_g(\Omega) =  \frac{\left(\frac{2\pi}{\Omega}\right)^{\frac{ n^2-1}{2}}e^{\Omega \hf(\rho_\ML)} }{\sqrt{\left| \det\left( \nabla^2 \hf_\ML \right) \right|}}
  \left( g(\rho_\ML) + O \left(\frac{1}{\Omega}\right)\right)
\end{equation}
Thus,  $ I_g(\Omega)= I_1 (\Omega)  \left( g(\rho_\ML) + O \left(\frac{1}{\Omega}\right)\right)$. With $g(\rho)=\tr{\rho A}\mathbb{P}_0(\rho)$   for any operator $A$, we get, assuming $\mathbb{P}_0(\rho_\ML)>0$:
\begin{equation}\label{eq:AsympExpanA}
\tfrac{\int_\DD \tr{A\rho} e^{\Omega \hf(\rho)} \mathbb{P}_{0}(\rho) d\rho }{\int_\DD e^{\Omega \hf(\rho)} \mathbb{P}_{0}(\rho) d\rho }= \tr{A \rho_\ML} + O \left(\frac{1}{\Omega}\right).
\end{equation}
 If $g(\rho_\ML)=0$, then $ I_g(\Omega)=  O \left(\frac{1}{\Omega}\right)$ and we have to use the next term in the asymptotic expansion. It appears that,
 when $g$ and its gradient vanish at $\rho_\ML$,  \cite[eq. 8.3.50 and 8.3.53]{BleistenHandelsmanBook} yield an explicit expression of this term with respect to $\nabla^2 g_\ML$, the Hessian of $g$ at $\rho_\ML$.   For $g(\rho)=h(\rho)\trr{(\rho-\rho_\ML) A}$ where $h$ is any scalar function of $\rho$ and $A$ is any Hermitian matrix of zero-trace, we have $\nabla^2 {g}_\ML (B) =2h(\rho_\ML) \tr{AB}A$ for any zero-trace Hermitian matrix $B$.  For such special form of $g$,  equations 8.3.50 and 8.3.53 of \cite{BleistenHandelsmanBook}  give:
\begin{multline}\label{eq:AsympExpanHg}
  I_{g}(\Omega)= -\left( \frac{h(\rho_\ML)}{\Omega}+ O \left(\frac{1}{\Omega^2}\right)\right) \ldots \\
 \ldots\left( \tfrac{\left(\frac{2\pi}{\Omega}\right)^{\frac{ n^2-1}{2}}e^{\Omega \hf(\rho_\ML)}\tr{A ~\nabla^2 {\hf}_\ML^{-1}(A)} }{\sqrt{\left| \det\left( \nabla^2 {\hf}_\ML\right) \right|}}\right)
  .
\end{multline}
Thus with $h=\mathbb{P}_0$, we get:
\begin{multline}\label{eq:AsympExpanSigA}
 \tfrac{\int_\DD \trr{A(\rho-\rho_\ML)} e^{\Omega \hf(\rho)} \mathbb{P}_{0}(\rho) d\rho }{\int_\DD e^{\Omega \hf(\rho)} \mathbb{P}_{0}(\rho) d\rho }
 \\= - \frac{\tr{A ~\nabla^2 {\hf}_\ML^{-1} (A)}}{\Omega} + O \left(\frac{1}{\Omega^2}\right)
 .
\end{multline}
Notice that  $\Omega \nabla^2 {\hf}$ corresponds to  the  Hessian of the restriction of  $f$ to  $\DD$. Some usual calculations show that this Hessian coincides with  $-\bR$
 where $\bR$ is defined in~\eqref{eq:FisherInfo} with  $\nabla f_\ML-\lambda_\ML I=0$ and $P_\ML=I$.  This explains the approximation~\eqref{eq:Avariance} since $A$ is of zero trace here and thus coincides with $A_\parallel$ because $\rho_\ML$  has full rank.

 When $\rho_\ML$ is rank deficient, then $\hf$ reaches its maximum on the boundary of $\DD$ and  the computations are more complicated. We only provide here the main steps  for $\rho_\ML$ of rank $n-1$. Moreover, we assume that $\nabla \hf_\ML \neq 0$.  The other  cases of rank between $1$ and $n-2$ can be conducted in a similar way and will be detailed in a forthcoming publication.  According to~\cite[eq. 8.3.10 and 8.3.11]{BleistenHandelsmanBook}, the first term of the asymptotic expansion  of $I_g(\Omega)$ coincides with the first term of the asymptotic expansion of a boundary  integral   localized on $\VV$, an open  small neighbourhood of $\rho_\ML$ on $\DD_{n-1}$:
 $$
J_g(\Omega)= \frac{1}{\Omega} \int_{\VV} g(\rho) \tfrac{\tr{\nabla \hf(\rho)  N(\rho)}}{\tr{\big(\nabla \hf(\rho)\big)^2}} e^{\Omega \hf(\rho)} d \Sigma
 $$
 where   the Hermitian operator $\nabla \hf$ is the gradient of $\hf$, the Hermitian operator  $N(\rho)$ corresponds to the unitary  normal  to $\DD_{n-1}$  at $\rho$ and  where $d\Sigma$ is the volume element on  $\VV$ considered as Riemannian submanifold of the Euclidian space of Hermitian matrices equipped with the Frobenius product. With
 $g_1(\rho)= g(\rho) \tfrac{\tr{\nabla \hf(\rho)  N(\rho)}}{\tr{\big(\nabla \hf(\rho)\big)^2}}$, we have to evaluate the following integral
$\int_{\VV} g_1(\rho)  e^{\Omega \hf(\rho)} d \Sigma$. Once again, we exploit~\cite[eq. 8.3.52]{BleistenHandelsmanBook} for this integral since $\hf$ restricted to $\VV$ reaches its maximum in the interior of $\VV$ with an invertible Hessian. We get:
$$
J_g(\Omega)
= \frac{1}{\Omega}\frac{K_\ML e^{\Omega \hf(\rho_\ML)} }{\Omega^{\frac{ n^2-2}{2}}}
  \left( g_1(\rho_\ML) + O \left(\frac{1}{\Omega}\right)\right),
$$
where $K_\ML$ is a positive constant, independent of $g_1$ and that can be expressed  via the  Hessian of $\hf$ restricted   to the Riemannian submanifold $\VV$.
 After some simple computations, we recover the asymptotic expansion~\eqref{eq:AsympExpanA}.

For $g(\rho)=h(\rho) \trr{(\rho-\rho_\ML) A} $ and  the corresponding $g_1(\rho)$,
we exploit~\cite[eq. 8.3.50 and 8.3.53]{BleistenHandelsmanBook}.  We get the following analogue
to~\eqref{eq:AsympExpanHg}:
\begin{multline*}
J_g(\Omega)  = -\left( \frac{h_1(\rho_\ML)}{\Omega  }+ O \left(\frac{1}{\Omega^2}\right)\right) \ldots \\
\left( \tfrac{ K_\ML e^{\Omega \hf(\rho_\ML)}  \tr{\widetilde{A} ~\widetilde{\nabla^2 \hf}_\ML^{-1} (\widetilde{A}) }}{\Omega^{\frac{ n^2}{2}}}\right)
\end{multline*}
with $h_1(\rho)= h(\rho) \tfrac{\tr{\nabla \hf(\rho)  N(\rho)}}{\tr{\big(\nabla \hf(\rho)\big)^2}}$.
Here, the Hermitian matrix $\widetilde A$ is given by the Hessian at $\rho_\ML$  of the restriction of  $\rho\mapsto \trr{(\rho-\rho_\ML) A}$ to the Riemannian submanifold $\VV$. Thus $\widetilde{A}$ is equal to the orthogonal projection onto the tangent space  at $\rho_\ML$ to the submanifold $\VV$. Similarly, $\widetilde{\nabla^2 \hf}_\ML$ corresponds to the Hessian at $\rho_\ML$  of  $\hf$ restricted to  the  submanifold  $\VV$.  With the  two  previous asymptotic expansions  for  the boundary integral $J_g(\Omega)$,  we  recover~\eqref{eq:AsympExpanSigA} with $A_\parallel$ instead of $A$ and $\widetilde{\nabla^2 \hf}_\ML$ instead of ${\nabla^2 \hf}_\ML$.  Some  additional calculations show that $\widetilde{A}$ corresponds to $A_\parallel$ and  $\Omega \widetilde{\nabla^2 \hf}_\ML$ to the opposite of $\bR$ defined in~\eqref{eq:FisherInfo}.

\end{document}